# Control of the Bright-Dark Exciton Splitting using Lamb Shift in a 2D Semiconductor

L. Ren (任磊)[1*], C. Robert[1*], M. Glazov[2*†], M. Semina[2], T. Amand[1], L. Lombez[1], D. Lagarde[1], T. Taniguchi[3], K. Watanabe[4] and X. Marie[1†]

[1]*Université de Toulouse, INSA-CNRS-UPS, LPCNO, 135 Av. Rangueil, 31077 Toulouse, France*
[2] *Ioffe Institute, 26 Polytechnicheskaya, 194021 Saint Petersburg, Russia*
[3]*International Center for Materials Nanoarchitectonics, National Institute for Materials Science, 1-1 Namiki, Tsukuba 305-00044, Japan*
[4]*Research Center for Functional Materials, National Institute for Materials Science, 1-1 Namiki, Tsukuba 305-00044, Japan*

*We have investigated the exciton fine structure in atomically thin $WSe_2$ -based van der Waals heterostructures where the density of optical modes at the location of the semiconductor monolayer can be tuned. The energy splitting Δ between the bright and dark exciton has been measured by photoluminescence spectroscopy. We demonstrate that Δ can be tuned by a few meV, as a result of a significant Lamb shift of the optically active exciton which arises from emission and absorption of virtual photons triggered by the vacuum fluctuations of the electromagnetic field. We also measured strong variations of the bright exciton radiative linewidth, as a result of the Purcell effect. All these experimental results illustrate the strong sensitivity of the excitons to local vacuum field. We found a very good agreement with a model that demonstrates the equivalence, for our system, of a classical electrodynamical transfer matrix formalism and quantum-electrodynamical approach. The bright-dark splitting control demonstrated here should apply to any semiconductor structures.*

\* These authors contribute equally to this work
† Corresponding authors : marie@insa-toulouse.fr, glazov@coherent.ioffe.ru

The exciton fine structure plays a crucial role for the light-matter coupling in semiconductor nanostructures [1]. The short range part of the electron-hole exchange interaction yields a splitting of the exciton states corresponding to different relative orientations of electron and hole spins [2–4]. As a result, the lowest energy exciton state is usually a dark state which is optically inactive. In (In)GaAs quantum wells or quantum dots this splitting between bright and dark exciton states is rather small, of the order of hundreds of μeV [5–8]. In 2D perovskites or 2D materials based on Transition Metal Dichalcogenides (TMDs), the dark exciton states can lie tens of meV below the bright ones as a result of larger exchange interaction due to very tightly bound excitons and specificities of the band structure [9–16]. In that case the exciton fine structure can have a dramatic impact on the emission yield of these new semiconductor nanostructures, even at room temperature [11,12].

So far one has considered that the splitting between bright and dark exciton in semiconductors is solely governed by the band structure and the amplitude of the exchange interaction between the electron and the hole. In this Letter we demonstrate that the coupling to light has also to be taken into account [17,18]. We evidence in our structure a clear tuning of the bright-dark exciton splitting as a result of a significant Lamb shift of the optically active exciton. This shift results from the emission and re-absorption of virtual photons, similarly to atomic systems [19–21]. In contrast, the dark exciton has an oscillator strength orders of magnitude smaller than the bright exciton one yielding a negligible energy shift due to the optical environment. As a consequence, the energy difference between the bright and the dark exciton varies with the characteristics of the electromagnetic field at the location of the semiconductor nanostructure. This can be achieved in a van der Waals heterostructure in the weak light-matter coupling regime as we demonstrate in this Letter or by tuning the distance between the semiconductor layer and an external mirror.

We have measured the variation of the bright-dark exciton splitting in a 2D semiconductor based on a $WSe_2$ monolayer (ML) encapsulated in hexagonal boron nitride (hBN). The control of the electromagnetic field distribution at the ML plane is simply achieved by changing the thickness of the hBN encapsulation layer, see Fig. 1(a). In a simplified picture, this is equivalent to a variation of the distance between the 2D layer and a mirror whose effective reflectivity is given by the stacking of different layers. The key advantage of this technique is linked to the role of the hBN encapsulation layer which yields narrow optical transitions approaching the homogenous exciton linewidth governed by radiative recombination [22–27]. We measured a variation of the bright-dark exciton splitting with the hBN thickness as large as 1.7 meV. This is a consequence of Lamb shifts that are orders of magnitude larger than the ones in atoms owing to the huge exciton oscillator strength in 2D semiconductors [28,29]. The key role of Quantum Electrodynamics (QED) in these structures is also attested by the simultaneous measurement of strong variations of the radiative linewidth, as a result of the Purcell effect [30]. Our measurements are in very good agreement with the calculated dependence of both the radiative linewidth and the bright-dark energy splitting using both transfer matrix techniques and QED approaches. We also uncover a relation between the Purcell and Lamb effects in TMD-based van der Waals heterostructures. We emphasize that the control of the bright-dark exciton splitting demonstrated here for a TMD ML is a general effect obtained in the weak exciton-photon coupling regime and should apply in principle to any semiconductor nanostructures [5,7,8,10].

*Samples and experimental set-ups*. The investigated samples are $WSe_2$ MLs encapsulated in hBN and deposited onto a 83 nm $SiO_2$/Si substrate using a dry-stamping technique [31]. Details on the samples can be found in the Supplemental Material [32]. The same $WSe_2$ ML is deposited on a hBN flake exhibiting different terraces and steps with hBN thicknesses $d$ measured during the sample fabrication by Atomic Force Microscopy, Fig. 1(a). The top hBN thickness does not play an important role here considering its small value, typically ~5-10 nm. Continuous wave (cw) photoluminescence (PL) experiments are performed at $T$=5 K using a He-Ne laser (633 nm). The typical excitation power is 5 µW corresponding to the linear regime of excitation. We also present the results of time-resolved photoluminescence experiments on a WSe$_2$ monolayer tuned to charge neutrality in a gated device [33]. In this case, the sample is excited with a Ti:Sa mode-locked laser (695 nm, ~1.5 ps pulse width, 80 MHz repetition rate). The PL kinetics are recorded by a synchro-scan streak camera with a typical time-resolution of 2 ps [25,34].

*Experimental results.* Figure 1(b) displays the cw PL spectrum for $d$=214 nm. In agreement with previous reports, the luminescence of the WSe$_2$ ML is dominated by the recombination of the neutral bright exciton ($X^0$) and the spin-forbidden dark exciton ($X^D$) with an energy splitting of $\Delta$=41 meV [11–16]. The optical selection rules dictate that the $X^D$ exciton is optically forbidden for in-plane polarized light but it can couple to $z$-polarized light [35]. Here, the light propagates mainly along $z$ (perpendicular to the ML plane) but we use a microscope objective with high numerical aperture (NA=0.82), yielding the detection of a fraction of $z$-polarized luminescence [14]. This explains the clear dark exciton $X^D$ PL line in Fig. 1(b). We also observe much smaller PL components, associated to the recombination of singlet ($X^{S-}$) and triplet ($X^{T-}$) negatively charged excitons and indirect exciton $X_I$ as already identified in many reports [36–38].

Figure 1(c) presents the time evolution of both bright ($X^0$) and dark ($X^D$) exciton luminescence following a picosecond excitation laser pulse in a WSe$_2$ monolayer with $d$= 290 nm. As already measured previously, the $X^0$ lifetime is very short, typically less than 2 ps as a result of the fast radiative recombination time of tightly bound excitons in TMDs [25,34]. In contrast, we measure a much longer PL decay time, ~ 800 ps, for the dark exciton $X^D$. This result tells us that the dark exciton oscillator strength is at least 3 orders of magnitude weaker than the bright exciton one [35]. As a consequence, we can assume that the Lamb shift of the dark exciton is negligible compared to the possible energy shift of the bright exciton linked to absorption/emission of virtual photons. Moreover, since the dark exciton modes are $z$-polarized, they do not experience any cavitylike effect in our structure.

First, we investigate the dependence of the $X^0$ luminescence linewidth as a function of $d$. Figure 2(a) displays the normalized PL spectra for $d$=101 and 186 nm. We observe a clear increase by more than a factor 2 (4.3 meV compared to 2 meV) in the luminescence linewidth (Full Width at Half Maximum, FWHM). As shown in Figure 3(a), $d$=101 nm and $d$=186 nm correspond to position of the ML at the node and the anti-node of the optical field intensity respectively in the cavitylike structure (calculations based on the transfer matrix method [39]). Thus the larger PL linewidth in Figure 2(a) for $d$=186 nm reflects the decrease of the radiative recombination time due to Purcell effect already observed in MoSe$_2$ MLs [25–27,40]. Figure 3(a) presents the variation of $X^0$ FWHM for eleven values of $d$, confirming the clear control of the linewidth due to the cavity effect. In order to reduce uncertainties, each value displayed in Fig. 3(a) is the average of about twenty measurements obtained at different points of the ML flake for a fixed $d$. The novelty here is the demonstration of the effect in WSe$_2$ monolayer. In contrast to MoSe$_2$, the bright exciton in WSe$_2$ monolayer lies *above* the dark exciton $X^D$, as shown in Fig. 1(b). The clear dependence of the bright exciton linewidth evidenced in Fig. 2(a) demonstrates that it is dominated by the radiative recombination and that the relaxation channel from cold bright exciton to the lower lying dark exciton plays a minor role.

Figures 2(b) and 3(b) present the key result of this Letter. We have measured the bright-dark exciton splitting Δ for the same samples and hBN thicknesses $d$ as the ones used for the investigation of the Purcell effect. Figure 2(b) displays for instance the PL spectra for $d$=132 and $d$=214 nm (the energy origin has been chosen at the dark exciton $X^D$ energy). We observe very clearly a variation $\delta E \approx 1.7$ meV of the splitting. The variation $\delta E$ of the bright-dark splitting as a function of $d$ is displayed in Fig. 3(b). Note that the splitting between $X^0$ and $X^D$ is $\Delta + \delta E$, choosing $\delta E = 0$ for $d$=100 nm, i.e. when the WSe$_2$ ML is at the node of the electric field in the cavitylike structure. We evidence a significant and oscillatory modulation of $\delta E$ as a function of the electromagnetic field amplitude. These results demonstrate that the energy difference between bright and dark excitons is not only controlled by electron-hole Coulomb exchange interaction and the semiconductor band structure parameters but the coupling to the electromagnetic field has also to be considered.

*Theory and discussion*. Both the exciton linewidth variation and the tuning of the bright-dark exciton energy splitting presented in Fig. 3 can be well understood on the basis of a model based on transfer matrix formalism and quantum-electrodynamical approaches (see Supplemental Information [32]). We have calculated the linear response functions (reflectivity, transmission and absorbance) of our stacking "top hBN layer/WSe$_2$ ML/bottom hBN layer/SiO$_2$/Si" with the measured 10 nm top hBN thickness, 83 nm SiO$_2$ thickness and crucially the variable bottom hBN thickness $d$. We used the following refractive indices: $n_{hBN}$ = 2.2, $n_{SiO2}$ = 1.46, $n_{Si}$ = 3.5 [39].

The full line in Fig. 3a is the calculated dependence of the bright exciton linewidth $\Gamma = \Gamma_0 + \Gamma_{nr}$, where $\Gamma_0$ and $\Gamma_{nr}$ are the radiative and non-radiative contributions respectively extracted from the calculated absorption spectrum. We notice a very good agreement between the measured and calculated dependence using the exciton radiative rate in vacuum $\Gamma_0^{vac} = 2$ meV and $\Gamma_{nr}$=0.6 meV. Note that the radiative decay rate is consistent with previous experimental and theoretical estimations of the recombination rate in WSe$_2$ monolayer where the cavity effect was not considered [34,41,42]. Remarkably the same parameters in the model also yield a very good description of the dependence of the exciton bright-dark splitting as a consequence of the Lamb shift, see the full line in Fig. 3(b).

In quantum electrodynamics, both the variation of the radiative decay rate and energy of the exciton stem from its coupling with vacuum fluctuations of electromagnetic field. Change in the bottom hBN thickness $d$ changes the local structure of electromagnetic modes in the system and, consequently, $\Gamma_0$ and $\delta E$. The analysis in SM [32] shows that these quantities can be also evaluated semiclassically, using the transfer matrix method and expressed via the electrodynamical Green's function. Compact analytical expressions can be derived neglecting the cap layer effect, in that case [25,32]:

$$\Gamma_0 + i\,\delta E = \Gamma_0^{vac}\bigl(1 + r_{bg}\bigr), \qquad (1)$$

where $r_{bg}$ is the complex reflection coefficient of a three-layer structure "hBN/SiO$_2$/Si". Thus the Purcell factor and Lamb shift are proportional to the real and imaginary parts of the substrate's reflection coefficient. It is instructive to consider an illustrative case of a simplified open cavity structure based on a WSe$_2$ ML lying at a distance $d'$ from a non-absorbing mirror characterized by a real reflection coefficient $r$ (inset in Figure 4) [29]. In that case $r_{bg} = r \exp(2iqd')$, $q = \omega/c$, $c$ is the speed of light and $\hbar\omega$ the exciton energy; the exciton radiative linewidth and the bright-dark splitting variation write simply [32]:

$$\Gamma_0 = \Gamma_0^{\text{vac}}[1 + r\cos(2qd')], \tag{2a}$$

$$\delta E = r\,\Gamma_0^{\text{vac}} \sin(2qd'). \tag{2b}$$

These simple expressions directly show why the two measurements are in quadrature in Fig. 3. In contrast to the linewidth which exhibits as expected minima and maxima at the nodes and anti-nodes respectively, the bright-dark splitting variation $\delta E$ is strictly zero for these two positions, in perfect agreement with Eq. (2b). Such a behavior is general and follows from the dispersion relations for the reflectivity in Eq. (1), see SM [32] for details. Figure 4 presents the calculated dependence of the bright-dark splitting energy variation for a $WSe_2$ monolayer in this simple open cavity composed of a mirror with a 100% reflection coefficient; we used here the same exciton radiative rate in vacuum $\Gamma_0^{\text{vac}} = 2$ meV as the one used in the calculated curves in Fig. 3. A variation of the bright-dark splitting as large as 4 meV due to the Lamb shift can be obtained. These huge variations result from Lamb shift orders of magnitude larger than the ones evidence in atomic systems [17–21,43].

It was recognized so far that the bright-dark exciton splitting $\Delta$ in TMD semiconductor monolayers includes three contributions [16,43]: $\Delta_3 = \Delta_{\text{exch.}} + \Delta_{\text{SO}} + \Delta_{\text{bind}}$, where $\Delta_{\text{exch.}}$ is the short range exciton exchange energy, $\Delta_{\text{SO}}$ is the conduction band spin–orbit splitting, $\Delta_{\text{bind}}$ is the difference between the binding energies of bright and dark excitons. The terms $\Delta_{\text{SO}}$ and $\Delta_{\text{bind}}$ are due to the specific band structure of TMD monolayers, whereas the exchange term is $\Delta_{\text{exch.}} \sim 10$ meV [38,45]. In Fig. 4, we show that the new contribution due to QED effect, $\delta E$, could be as large as 4 meV, i.e. ~ 40% of the exchange term resulting in $\Delta = \Delta_3 + \delta E$. We can anticipate that similar effects should occur in other TMD monolayers, including $WS_2$ and $MoS_2$ [46–49]. For $MoSe_2$ monolayers, the impact of the optical environment could be even more interesting since the dark exciton lies slightly above the bright one with reported values of $\Delta = -1.5$ meV [16,50], *i.e.* of the order of the energy shifts we have evidenced in $WSe_2$ MLs, Fig. 3(b). This means that a proper engineering of the vacuum field quantum fluctuations should reverse the bright-dark exciton ordering in $MoSe_2$ MLs. We emphasize that this should occur in *the weak coupling regime*, in contrast to the change of bright-dark ordering evidenced for exciton-polaritons *in the strong coupling regime* for a $WSe_2$ monolayer placed inside a high finesse optical cavity [51]. In GaAs quantum wells, the free exciton oscillator strength is typically ~10 times weaker than the one in TMD monolayers [50]; thus Eq. (2b) predicts a typical variation of the bright-dark splitting of the order of ~100 μeV (using *r*=1). Remarkably this value is similar to the bright-dark splitting due to the exchange interaction measured in GaAs/AlGaAs quantum wells [5,6]. For 2D perovskites significant variations of the exciton bright-dark splitting are also expected [9].
Finally we can note that the electromagnetic quantum fluctuations could also impact long-range exchange interaction in the exciton, as theoretically predicted both in bulk and 2D semiconductors [1,53,54].

In conclusion we have shown that not only the exciton radiative lifetime is controlled by the optical environment in a 2D semiconductor but the bright-dark exciton splitting can also be modified by the potential induced by quantum fluctuations of the electromagnetic field. All these results show that the excitons in semiconductor nanostructures are very sensitive probes of the local vacuum field.

*Acknowledgments:* This work was supported by the French Agence Nationale de la Recherche under the program ESR/EquipEx+ (grant number ANR-21-ESRE- 0025) and the ANR projects ATOEMS, Sizmo-2D and Magicvalley.

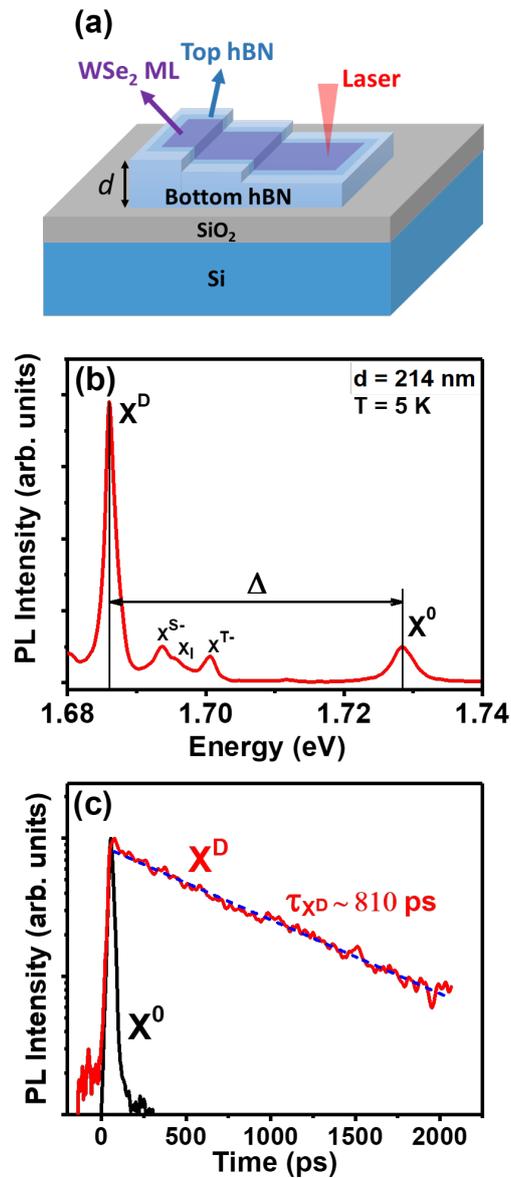

*Figure 1. (a) Schematics of the investigated hBN-encapsulated WSe$_2$ monolayer ; (b) cw photoluminescence spectrum of the WSe$_2$ monolayer (d=214 nm) showing mainly the emission of the neutral bright ($X^0$) and dark exciton ($X^D$) at T = 5 K, see text. The energy difference between $X^0$ and $X^D$ is denoted Δ. (c) Normalized photoluminescence intensity as a function of time for the neutral bright ($X^0$, black solid line) and dark exciton ($X^D$, red solid line), d=290 nm. The blue dashed line corresponds to a mono-exponential fit of the decay time $\tau_{X^D}$.*

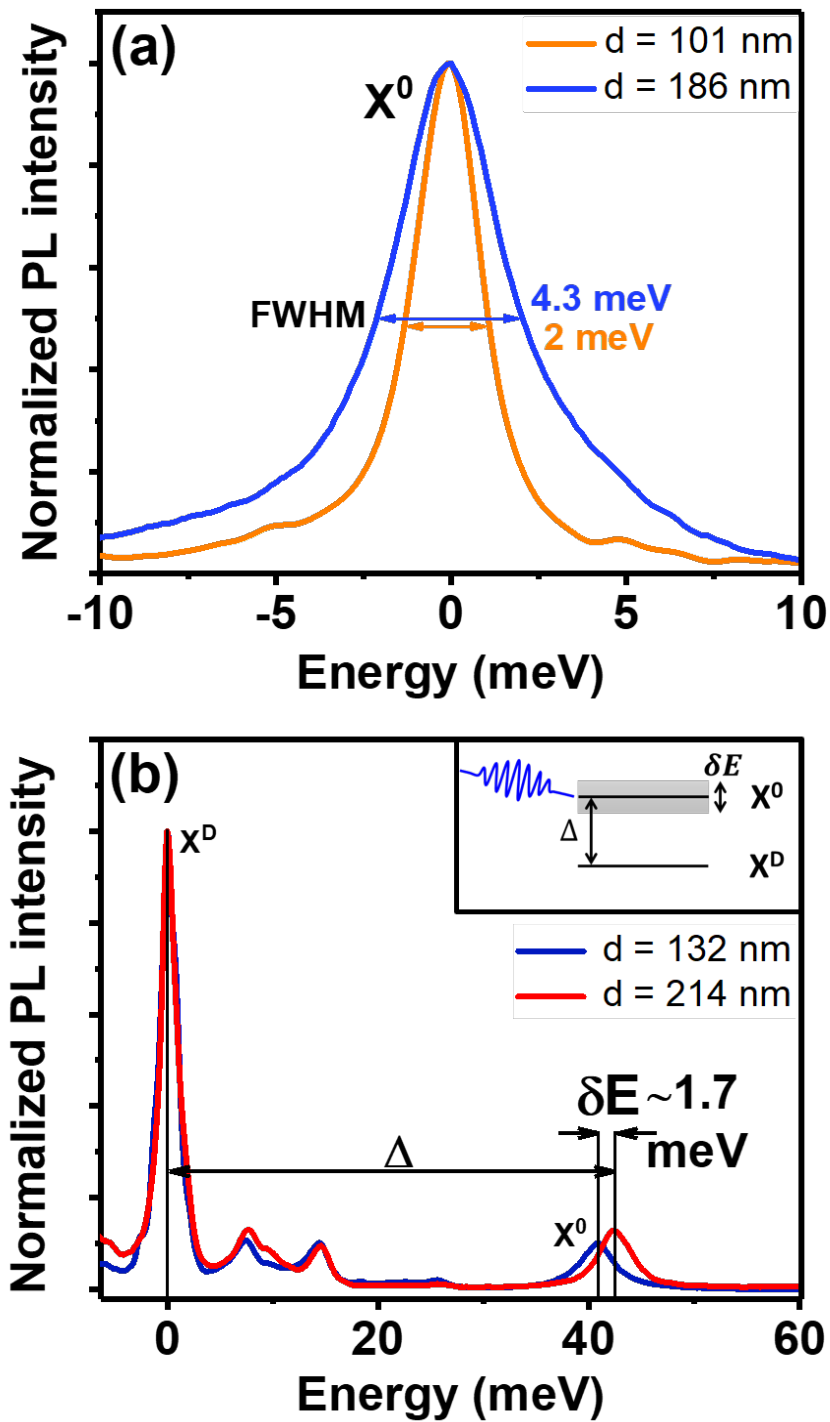

*Figure 2. (a) Normalized cw PL intensity of the neutral exciton for d = 101 nm and d = 186 nm. In order to compare the linewidths, the origin of the energy axis is taken at the PL peak. The double arrow lines indicate the FWHM linewidths. (b) Normalized PL spectra for d=132 nm and d=214 nm. The energy axis is taken at the $X^D$ PL peak. Inset: schematics of the transition energy shift of $X^0$ (grey shadow) due to Lamb shift.*

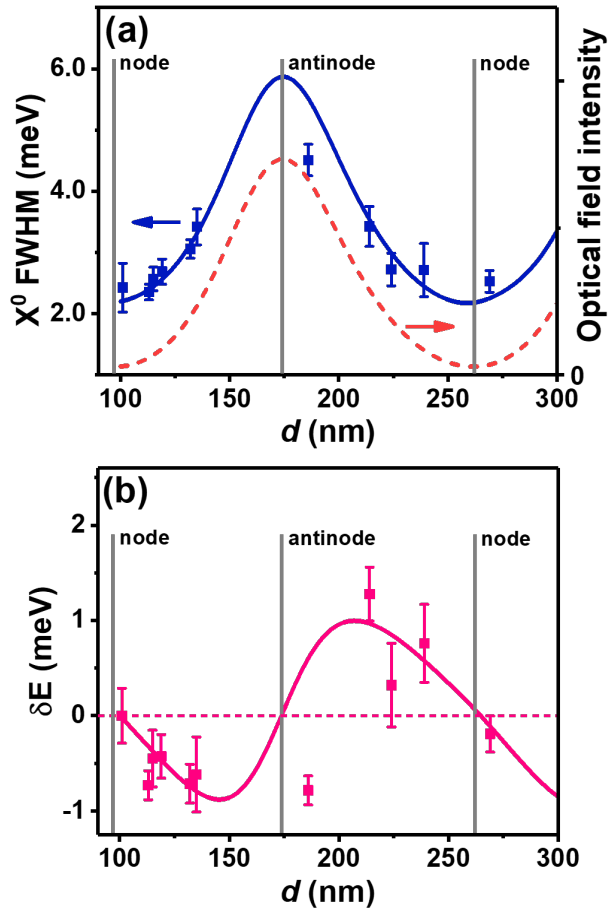

*Figure 3. (a) Measured (blue symbols) and calculated (blue solid line) neutral bright exciton linewidth as function of the bottom hBN thickness d. The red dashed curve is the calculated optical field intensity at the monolayer plane. The thickness where the ML is located at the nodes and antinodes are indicated by the vertical grey bars. (b) Measured (pink symbols) and calculated (pink solid line) variation δE of the bright-dark exciton splitting as a function of d.*

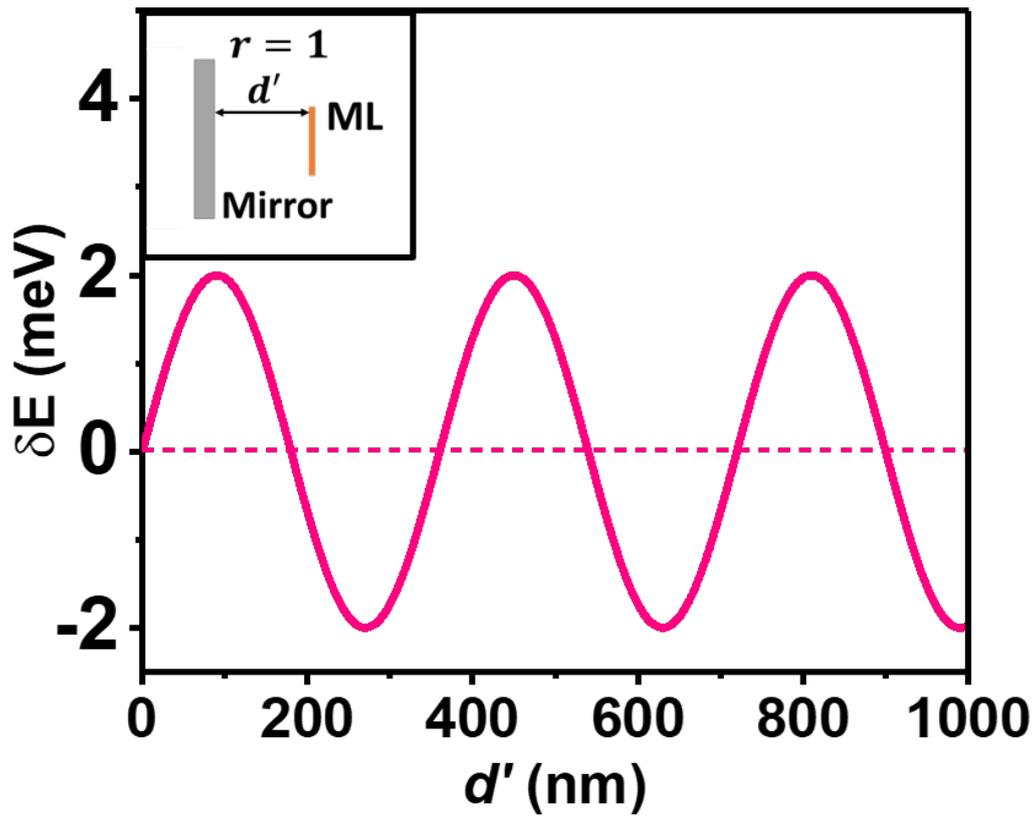

*Figure 4. Calculated variation δE of the bright-dark exciton splitting as a function of the distance d' between the mirror (with reflection coefficient $r = 1$) and the WSe2 monolayer. Inset: schematics of the simple configuration where the monolayer is at a distance d' from the mirror.*

# Supplemental Material:

# Control of the Bright-Dark Exciton Splitting using Lamb Shift in a 2D Semiconductor

L. Ren(任磊)[1*], C. Robert[1*], M. Glazov[2*†], M. Semina[2], T. Amand[1], L. Lombez[1], D. Lagarde[1], T. Taniguchi[3], K. Watanabe[4] and X. Marie[1†]

[1]*Université de Toulouse, INSA-CNRS-UPS, LPCNO, 135 Av. Rangueil, 31077 Toulouse, France*
[2] *Ioffe Institute, 26 Polytechnicheskaya, 194021 Saint Petersburg, Russia*
[3]*International Center for Materials Nanoarchitectonics, National Institute for Materials Science, 1-1 Namiki, Tsukuba 305-00044, Japan*
[4]*Research Center for Functional Materials, National Institute for Materials Science, 1-1 Namiki, Tsukuba 305-00044, Japan*

**SI. SAMPLES DESCRIPTION**

Van der Waals heterostructures (top hBN/WSe$_2$/bottom hBN) are fabricated by exfoliating WSe$_2$ monolayers and high quality hBN layers. The WSe$_2$ bulk material is purchased at 2D Semiconductors and the hBN crystals are provided by NIMS (Japan). All the layers are successively deposited on to a 83 nm SiO$_2$/Si substrate by a dry-stamping technique [1]. The bottom hBN thicknesses *d* are measured by Atomic Force Microscopy (AFM, CSI Instrument). We list below the four samples used in the main text (Table S1), including two "rainbow samples" where the same WSe$_2$ ML is deposited on a hBN flake exhibiting different terraces and steps. Figure S1 shows an optical microscope image of Sample IV.

| Sample | Bottom hBN thickness d (nm) |
|---|---|
| Table S1 Samples characterized with different bottom hBN thickness. | |
| Sample I | 101 |
| Sample II (rainbow) | 113 |
| | 115 |
| | 119 |
| | 132 |
| | 135 |
| Sample III | 186 |
| Sample IV (rainbow) | 214 |
| | 224 |
| | 239 |
| | 269 |

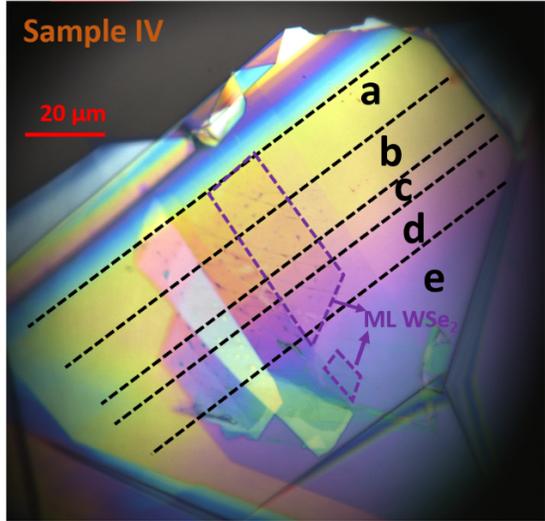

*Figure S1 Optical microscope image of Sample IV showing different bottom hBN thicknesses: a (d=210 nm), b (d=214 nm), c (d=224 nm), d (d=239 nm) and e (d=269 nm). The black dashed lines separate different terraces and the WSe$_2$ monolayer is indicated by the purple dashed line.*

## SII. CALCULATION OF THE EXCITON LAMB SHIFT AND PURCELL EFFECT

We consider a TMD ML in van der Waals heterostructures described in S1. [2]. Below we calculate the excitonic Lamb shift and radiative decay rate with two approaches: the classical electrodynamical Green's function approach (transfer matrix method) and quantum electrodynamical (QED) approach. We demonstrate that these approaches are equivalent for our problem.

In the classical electrodynamical Green's function approach we calculate the eigenmodes of electromagnetic field in the presence of a monolayer (exciton-polaritons in terms of bulk crystals) and extract the polariton energy and its damping from the elementary response functions, e.g., from the absorption [2,3,10]. In the QED approach we quantize the electromagnetic modes in the structure and calculate exciton energy renormalization due to coupling with the quantized modes using the second-order perturbation theory.

To illustrate that electrodynamical Green's function approach and quantum-electrodynamical approaches give the same results we consider a simplified setting of the problem with the monolayer being at a distance $d'$ from a mirror (barrier) characterized by the reflection coefficient $r$ and transmission coefficient $t$ (inset of Figure 4 in the main text). We assume that the mirror is non-absorbing.

## SII.1 GREEN'S FUNCTION FOR 1D ELECTROMAGNETIC WAVES

Our system is translationally invariant in a 2D monolayer plane ($xy$) and inhomogeneous along the $z$ axis. Let us consider the 1D Maxwell equation for the in-plane electric field

component $E = E_x$ or $E_y$ [3]

$$\frac{d^2E}{dz^2} + q^2\varepsilon(z)E = -4\pi q^2 P_{exc}(z) \tag{1}$$

Here $q = \omega/c$, $\varepsilon$ is the dielectric constant of the system and $P_{exc}(z)$ is the polarization induced by the exciton. We introduce the electrodynamical Green's function $G_\omega(z, z')$ as a solution of Eq. (1) with the $\delta$-function source:

$$\left[\frac{d^2}{dz^2} + q^2\varepsilon(z)\right] G_\omega(z, z') = \delta(z - z') \tag{2}$$

In a vacuum where $\varepsilon(z) = 1$ the Green's function reads

$$G_\omega(z, z') = \frac{\exp(iq|z - z'|)}{2iq} \tag{3}$$

In the presence of a barrier the Green's function can be also readily found by matching the solutions. To that end we assume that the barrier is at $z = 0$ and the monolayer is at $z \equiv z_{ML} = -d'$. For the calculations of the Purcell factor and Lamb shift presented in the following we need the Green's function at the monolayer only ($z = z' = z_{ML}$) but we present more general solution with $z' = z_{ML}$ and arbitrary $z < 0$:

$$G_\omega(z, -d') = \frac{1}{2iq}\begin{cases} e^{iq(z+d')} + re^{2iqd' - iq(z+d')}, & -d' < z < 0, \\ e^{-iq(z+d')}(1 + re^{2iqd'}), & z < -d' \end{cases} \tag{4}$$

It can be checked by substitution that the function (4) satisfies the Maxwell's equation (2), it is continuous at $z = z_{ML}$ and its derivative has a jump by 1 at $z = z' = z_{ML}$. Moreover, it satisfies the boundary condition at the barrier: The reflected wave amplitude is $r$ times the incident wave amplitude.

It is instructive to derive Eqs. (3) and (4) using the normal mode decomposition. In the free space the normal modes are the plane waves

$$E_{k_z}(z) = e^{ik_z z}, \tag{5}$$

where we assume the normalization length to be unity and

$$G_\omega(z, z') = \sum_{k_z} \frac{E_{k_z}(z) E^*_{k_z}(z')}{q^2 - k_z^2 + i0}, \quad \sum_{k_z} \cdots = \int_{-\infty}^{\infty} \frac{dk_z}{2\pi} \cdots. \tag{6}$$

The summation can be efficiently carried out using the residue theorem and taking into account that

$$q^2 - k_z^2 + i0 = (q - k_z + i0)(q + k_z + i0)$$

The pole with $k_z = q + i0$ should be used for $z - z' > 0$ and the pole with $k_z = -q - i0$ for $z - z' < 0$ as a result we have Eq. (3).

The calculation in the presence of the barrier is performed in the following way. Making use of the expressions presented in the supplement to Ref. [2] (see also Ref. [4]) we obtain for the fields

$$E^{(1)}_{k_z}(z) = \begin{cases} e^{ik_z z} + r e^{-ik_z z}, & z < 0, \\ t e^{ik_z z}, & z > 0, \end{cases} \quad (7a)$$

$$E^{(2)}_{k_z}(z) = \begin{cases} t' e^{-ik_z z}, & z < 0, \\ e^{-ik_z z} + r' e^{ik_z z}, & z > 0. \end{cases} \quad (7b)$$

In both cases we assume that $k_z > 0$, the wave (1) corresponds the field incident from the $z = -\infty$ and the wave (2) corresponds to the field incident from $z = +\infty$. The time-reversal symmetry dictates the following requirement for the relation between $t$, $t'$, $r$ and $r'$:[1]

$$t' = t, \quad r' = -r^* \frac{t}{t^*}, \quad \text{or } t^* r' + r^* t = 0 \quad (8)$$

Hence, $|r| = |r'|$ and, in the absence of losses, $|t|^2 + |r|^2 = 1$. As a result the modes are properly normalized and orthogonal, see Ref. [2,4] for details. Let us find the Green's function at $z = z' = -d'$ explicitly using the normal mode decomposition [cf. Eq. (6)]

$$G_\omega(z, z') = \sum_{i=1,2} \int_0^\infty \frac{dk_z}{2\pi} \frac{E^{(i)}_{k_z}(z) E^{(i)*}_{k_z}(z')}{q^2 - k_z^2 + i0}, \quad (9)$$

namely

$$G_\omega(z_{ML}, z_{ML}) = \int_0^\infty \frac{dk_z}{2\pi} \frac{|e^{-ik_z d} + r e^{ik_z d}|^2}{q^2 - k_z^2 + i0} + \int_0^\infty \frac{|t'|^2}{q^2 - k_z^2 + i0} \frac{dk_z}{2\pi}, \quad (10)$$

where the first term comes from the solution (1) and the second term comes from the solution (2) in Eq. (7).

We now show that Eq. (10) equivalent to Eq. (4). Note that the following relations are fulfilled for any non-absorbing barrier:

$$t'(k_z) = t(k_z), \quad r(k_z) = r^*(-k_z). \quad (11)$$

Both these relations are a consequence of the time-reversal symmetry. As a result, Eq. (10) can be recast as

$$G_\omega(z_{ML}, z_{ML}) = \frac{1}{2} \int_{-\infty}^\infty \frac{dk_z}{2\pi} \frac{2 + r e^{2ik_z d} + r^* e^{-2ik_z d}}{q^2 - k_z^2 + i0}, \quad (12)$$

where we made use of the fact that $|t|^2 + |r|^2 = 1$ and that the subintegral expression is an even function of $k_z$. Integration in Eq. (12) can be performed via the residue theorem. To that end we note that all poles of $r(k_z)$ in the complex plane have negative imaginary parts. It is a result of the causality.[2] Hence, while integrating the term $r\exp(2ik_z d')$ we close the contour in the upper half-plane, $\text{Im} k_z > 0$. In this way, the Jordan's lemma is fulfilled, and we need to take into account the only pole of the denominator with $k_z = q + i0$ and disregards the poles of $r(k_z)$. While integrating the term with $r^* \exp(-2ik_z d')$ we close the contour in the lower half-plane, $\text{Im} k_z < 0$. Again, the Jordan's lemma is fulfilled, and we need to take into account the only pole of the denominator with $k_z = -q - i0$. Making use again of Eq. (11) we arrive at

$$G_\omega(z_{ML}, z_{ML}) = \frac{1 + r\exp(2iqd')}{2iq}, \tag{13}$$

in full agreement with Eq.(4). The derived Green's function allows us to evaluate the Purcell effect and the Lamb shift semiclassically, see SII.3. Before doing so, let us present the results of the quantum-mechanical calculation.

---

[1] We recall that the transfer matrix through a barrier takes the form $\hat{T} = \begin{bmatrix} 1/t^* & -r^*/t^* \\ -r/t & 1/t \end{bmatrix}$

### SII.2 QUANTUM-MECHANICAL CALCULATION

We use standard QED approach to calculate the shift of the exciton energy within the second-order perturbation theory in the light-matter coupling. The renormalization of the energy results from the process of the virtual emission of a photon (exciton virtually recombines) and re-absorption of the same photon. Corresponding expression for the energy shift reads [5,6]

$$\delta E_{\text{exc}} = \sum_j \frac{|V_j|^2}{E_{\text{exc}} - \hbar\omega_j + i0}. \tag{14}$$

Here $j$ enumerates the photon states ($j = k_z$ in the case of a free space, or $j = (k_z, i)$ with $k_z > 0$ and $i = (1,2)$ in the presence of a barrier), $V_j$ is the matrix element of the light-matter interaction, $E_{\text{exc}}$ is the bare exciton energy, $\omega_j$ is the photon dispersion. We have added $i0$ in the denominator in accordance with the general prescriptions of the scattering theory to ensure causality, imaginary part of $\delta E_{\text{exc}}$ gives the exciton radiative broadening (with the negative sign):

$$\Gamma_0 = -\text{Im}\,\delta E_{\text{exc}}. \tag{15}$$

Our system consists of a monolayer semiconductor sandwiched between hBN layers and deposited on the SiO₂/Si substrate. Such a structure "hBN/SiO₂/Si" is highly non-resonant, and we need to take into account all modes of electromagnetic field propagating in the structure. Thus, in the case of interest, $V_j$ can be written as

$$V_j = -\sqrt{2\pi\hbar\omega_j}\, D_{\text{exc}} E_{k_z}^{(i)}(z_{ML}), \tag{16}$$

Where $D_{\text{exc}}$ is the exciton microscopic dipole moment, $E_{k_z}^{(i)}(z_{ML})$ are the eigenmodes of the field at the position of the monolayer, Eq. (7), and the factor $\sqrt{2\pi\hbar\omega_j}$ results from the field normalization per photon. Equation (14) provides the shift of the excited state of the crystal, namelty, of the exciton. However, the process of the photon emission and corresponding re-absorption also affects the energy of the ground state of the crystal as

$$\delta E_0 = \sum_j \frac{|V_j|^2}{-E_{exc} - \hbar\omega_j + i0}. \tag{17}$$

---

[2] This fact can be illustrated by a simple example of a barriers considered in Ref. [2], Eqs. (S36), (S37): Their denominators as functions of $k_z$ read $D(k_z) = 1 - \rho\exp(ik_z l)$ where $\rho$ is the product of corresponding reflection coefficients and $l$ is the effective length. The zeros of $D(k_z)$ take the form $k_z^{(n)} = il^{-1}\ln\rho + 2\pi n/l$ with $n \in \mathbb{Z}$ and, since $|\rho| < 1$ and $l > 0$, $\text{Im}\{k_z^{(n)}\} < 0$.

The observed energy shift of the bright exciton line is given by

$$\delta E = \delta E_{exc} - \delta E_0 = 2 E_{exc} \sum_j \frac{|V_j|^2}{E_{exc}^2 - |\hbar\omega_j|^2 + i0} = 4\pi\hbar\, E_{exc}|D_{exc}|^2 \sum_{\substack{k_z>0 \\ i=1,2}} \frac{c|k_z|\left|E_{k_z}^{(i)}(z_{ML})\right|^2}{E_{exc}^2 - (\hbar c k_z)^2 + i0}. \tag{18}$$

The real part of Eq. (18) diverges. This divergence is known from the quantum electrodynamics and related to the renormalization of the free electron mass. The simplest way to remove the divergence is, following Refs. [5,8], to remove the contribution $\delta m_{QED} v^2/2$ from the energy.[3] As a result we have

$$\delta E = 4\pi \left|\frac{E_{exc} D_{exc}}{\hbar c}\right|^2 G_{\frac{E_{exc}}{\hbar}}(z_{ML}, z_{ML}), \tag{19}$$

where in the last equality we introduced the Green's function of electrodynamical problem, Eqs. (9) and (12).

### SII.3 COMPARISON OF THE RESULTS

Let us now explicitly check that both approaches give the same result. In the classical electrodynamical treatment we need to solve Eq. (1) together with the material equation for the exciton polarization $P_{exc}$ [2,3,8]

$$(E_{exc} - \hbar\omega) P_{exc}(z) = |D_{exc}|^2 E(z_{ML}) \delta(z - z_{ML}). \tag{20}$$

Substituting Eq. (20) into Eq. (1) and solving it using the Green's function as

$$E(z) = -4\pi \left(\frac{\omega}{c}\right)^2 \frac{|D_{exc}|^2}{E_{exc} - \hbar\omega} E(z_{ML}) G_\omega(z, z_{ML}),$$

we arrive, setting $z = z_{ML}$, at the following self-consistency equation

---

[3] The renormalization can be performed as follows. We consider a shift of a given electronic level s as a result of its coupling with all the levels, cf. Eq. (14), and subtract the contribution related to the diagonal matrix element of the velocity in this state [5,8]:

$$\delta E_s = 2\pi \sum_{j,s'} \frac{\hbar\omega_j |D_{s's}|^2 |E_j|^2}{E_s - E_{s'} - \hbar\omega + i0} \to 2\pi \sum_{j,s'} \left\{ \frac{\hbar\omega_j |D_{s's}|^2 |E_j|^2}{E_s - E_{s'} - \hbar\omega + i0} - \frac{\hbar\omega_j |D_{s's}|^2 |E_j|^2}{-\hbar\omega + i0} \right\} = 2\pi \sum_{j,s'} \frac{(E_s - E_{s'})|D_{s's}|^2 |E_j|^2}{E_s - E_{s'} - \hbar\omega + i0}$$

$$\hbar\omega - E_{exc} = 4\pi \left|\frac{E_{exc}D_{exc}}{\hbar c}\right|^2 G_{\frac{E_{exc}}{\hbar}}(z_{ML}, z_{ML}), \tag{21}$$

where we replaced $\omega$ by $E_{exc}/\hbar$ in the right-hand side taking into account that the energy shifts are small [10,11] (we operate in a weak coupling regime). This calculation is fully equivalent to the transfer matrix calculation. It is seen that that Eq. (21) fully agrees with Eq. (19) derived above from the QED.

In free space for the renormalization of the energy and radiative damping we have:

$$\delta E^{vac} = 0, \tag{22a}$$

$$\Gamma_0^{vac} = 2\pi \frac{E_{exc}|D_{exc}|^2}{\hbar c}. \tag{22b}$$

The fact that $\delta E^{vac} = 0$ corresponds to the regularization procedure we used where all "remaining" quantum electrodynamical contributions are already included in the transition energy. In the presence of the barrier the results change as

$$\delta E = \Gamma_0^{vac} r \sin(2qd'), \tag{23a}$$

$$\Gamma_0 = \Gamma_0^{vac}(1 + r \cos(2qd')). \tag{23b}$$

The fact that $\delta E$ vanishes at the nodes and antinodes of the electromagnetic field, where $2k_z d' = n\pi$ with $n \in \mathbb{N}$, can be understood from simple qualitative arguments: According to Eq. (14) the energy shift of the exciton due to the coupling with the photon with the frequency $\omega_j$ is an odd function of the detuning $E_{exc} - \hbar\omega_j$, $\delta E \propto (E_{exc} - \hbar\omega_j)^{-1}$. For the case of the antinode or node of electromagnetic field at the monolayer, $k_z d' = n\pi/2$, the photon density of states is approximately symmetric with respect to $E_{exc}$ and the total shift vanishes. If the condition $k_z d' = n\pi/2$ is not fulfilled the photon density of states is no longer symmetric with respect to $E_{exc}$ and the Lamb shift appears as illustrated in Figure S2.

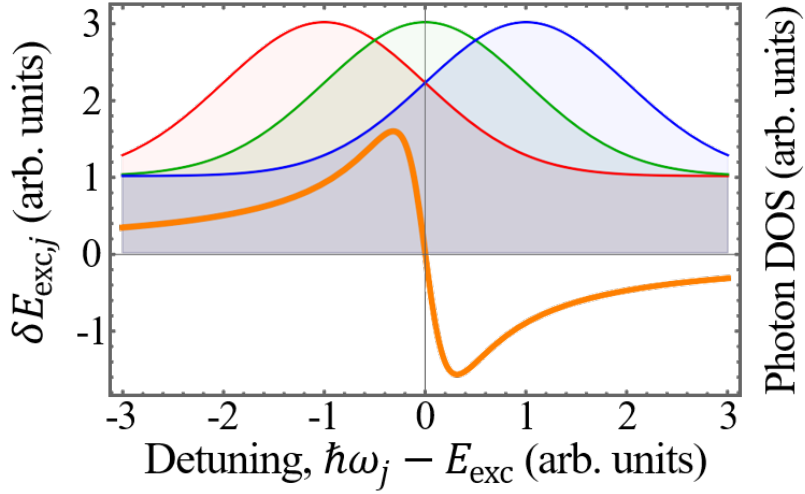

**Figure S2** *Illustration of the cancellation of the Lamb shift at the "resonant" case of the antinode of the field. Orange dispersive curve shows the contribution of the photon mode with the frequency $\omega_j$ to the Lamb shift, Eq. (14). Red, green, and blue peaked functions illustrate the photon density of states for the detuned cases (red and blue) and resonant, $k_z d' = n\pi/2$ (green) case.*

## SII.4 KRAMERS-KRONIG-LIKE RELATION BETWEEN THE LAMB SHIFT AND PURCELL EFFECT

Generally, in the structure with an arbitrary barrier the electrodynamical treatment shows that

$$\delta E = \Gamma_0 \text{Im}\{r\}, \quad (24a)$$

$$\Gamma_0 = \Gamma_0^{\text{vac}}(1 + \text{Re}\{r\}), \quad (24b)$$

where $r \equiv r(E_{\text{exc}}/\hbar)$ is the reflection coefficient of the barrier taken at the exciton resonance frequency. There is a deep connection between $\delta E$ and $\Gamma_0$ based on the causality principle. It follows that the real and imaginary parts of the reflection coefficient are interrelated by the Kramers-Kronig type relation [12,13]

$$\text{Im}\{r(\omega)\} = -\frac{2\omega}{\pi} v.p. \int_0^\infty \frac{\text{Re}\{r(\omega')\}}{\omega'^2 - \omega^2} d\omega', \quad \text{Re}\{r(\omega)\} = -\frac{2\omega}{\pi} v.p. \int_0^\infty \frac{\text{Im}\{r(\omega')\}}{\omega'^2 - \omega^2} d\omega'. \quad (25)$$

Note that the real and imaginary parts of the electrodynamical Green's function $G_\omega(z, z')$ obeys the same relations, see Ref. [13] for details. To demonstrate this relations we plotted in Figure S3 the reflection coefficient of a typical hBN/SiO$_2$/Si substrate calculated by the transfer matrix method [2,14] and, for the imaginary part, using the relation (25). A very good agreement (within the numerical accuracy) is seen. It also shows a phase-shift of a $\pi/2$ between the Purcell enhancement and the Lamb shift as a function of the hBN thickness.

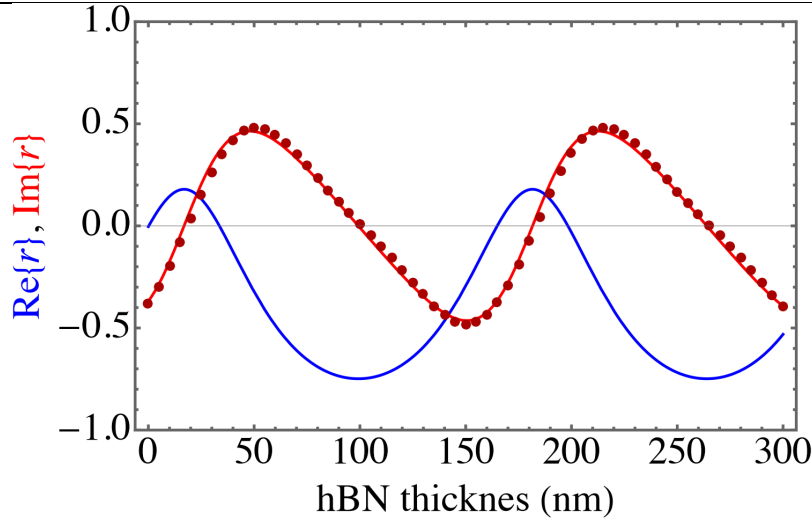

**Figure S3** Background reflectivity of the hBN/SiO$_2$/Si van der Waals heterostructure as a function of the hBN thickness at the WSe$_2$ exciton energy $E_{exc}$ = 1.71 eV. SiO$_2$ thickness is 83 nm. Blue and red curves show the real part of the reflection coefficient found using the transfer matrix method, Refs. [2,14] dark red dots show Im{r}, calculated after Eq. (25) via Re{r}.

In this work we have studied the Lamb shift in a structure which does not support well-confined (along z-axis) optical modes. It is instructive to compare these results with the case of a ML semiconductor placed in a microcavity with highly reflective mirrors [7]. Such structure possesses well-defined optical modes with low damping. As a result, the light-matter interaction occurs mainly with a single optical mode whose frequency $\omega_0$ is close to the exciton resonant frequency $E_{exc}/\hbar$. In this case it is sufficient to keep only one, resonant, term in Eq. (14) in the QED expression for the Lamb shift. In this way we recover standard expressions for the energy shift of an exciton resonance due to the polariton effect (in the weak coupling regime):

$$\delta E = \frac{|V_0|^2}{E_{exc} - \hbar\omega_0}.$$

We stress that the crucial difference between the microcavity structure and our structure of a ML on top of the hBN/SiO$_2$/Si substrate is the absence of a single optical mode that resonantly couples with exciton making QED calculation more involved.